\documentclass[11pt, letterpaper]{article}

\usepackage{graphicx}      
\usepackage{amsmath}
\usepackage{setspace}
\usepackage{amssymb}
\usepackage{epstopdf}
\usepackage{multirow}
\usepackage{algorithm}
\usepackage{algpseudocode}
\usepackage{color}
\usepackage[letterpaper,left=1in,right=1in,top=1in,bottom=1in]{geometry}

\usepackage{titlesec}
\newtheorem{remark}{Remark}
\newtheorem{proposition}{Proposition}
\newtheorem{theorem}{Theorem}

\newtheorem{assumption}{Assumption}
\newtheorem{definition}{Definition}

\newcommand{\RNum}[1]{\uppercase\expandafter{\romannumeral #1\relax}}
\newcommand{\rNum}[1]{\lowercase\expandafter{\romannumeral #1\relax}}

\newcommand{\m}{\mathbb}

\title{\Large \bf Robust MPC with Zone Tracking}
\author{
	\centerline{\normalsize Zhiyinan Huang$^{a}$, 
		\ Jinfeng Liu$^{a,}$\thanks{Corresponding author: J. Liu. Email: jinfeng@ualberta.ca.},\ Biao Huang$^{a}$}
	\vspace{5mm}\\
	\centerline{\small $^{a}$Department of Chemical \& Materials Engineering, University of Alberta,}\\
	\centerline{\small Edmonton, AB T6G 1H9, Canada}\\
	}
\allowdisplaybreaks
\begin{document}
	\date{}
	\maketitle
	\setstretch{1.39}
	
	\begin{abstract}                
		We propose a robust nonlinear model predictive control design with generalized zone tracking (ZMPC) in this work. The proposed ZMPC has guaranteed convergence into the target zone in the presence of bounded disturbance. The proposed approach achieves this by modifying the actual target zone such that the effect of disturbances is rejected. A control invariant set (CIS) inside the modified target zone is used as the terminal set, which ensures the closed-loop stability of the proposed controller. Detailed closed-loop stability analysis is presented. Simulation studies based on a continuous stirred tank reactor (CSTR) are performed to validate the effectiveness of the proposed ZMPC.
	\end{abstract}
	
	

	\section{Introduction}
	Due to the development in computational hardware, nonlinear model predictive control (MPC) is considered to be a promising advanced control strategy and has attracted high research interests over the past decades. Inspired by the linear quadratic regulator (LQR), MPC has the ability to solve nonlinear optimization problems while considering constraints simultaneously. However unlike LQR problems that can be solved explicitly for infinite horizon, the same cannot be achieved for MPC due to the presence of constraints. Instead, MPC is solved in a receding horizon manner \cite{meadows_receding_1995, rawlings_tutorial_1999}, where the optimization problem is solved for a finite control horizon at each sampling instant with only the very first control policy applied to the system. 
	Conventional MPC tracks a referencing trajectory, which is oftentimes an optimal steady-state operating point \cite{rawlings_tutorial_1999}. This approach, however, may limit the flexibility of the controller and sacrifice the process performance during transient operations. One popular extension of the conventional MPC is economic MPC (EMPC), which considers a general economic objective directly in dynamic optimizations \cite{rawlings_fundamentals_2012}. Improved transient economic performance has been observed in EMPC compared to MPC \cite{diehl_lyapunov_2011, ellis_tutorial_2014}. 
	
	Another approach that helps to improve the flexibility of conventional MPC is zone MPC (ZMPC). ZMPC relaxes the set-point-tracking objective to a zone-tracking one, which aims to drive the system into a bounded set \cite{scokaert_feasibility_1999, askari_stability_2017}. More degrees of freedom are provided by ZMPC, which are beneficial for handling multiple control objectives (e.g. tracking and economic) simultaneously. Furthermore, ZMPC is potentially more robust in the presence of noise or disturbance. Zone-tracking is also a natural objective that rises in many real-world problems. For example, in agricultural practice, maintaining the soil moisture in a certain range is often sufficient and more practical rather than attempting to maintain it at a particular level. Various applications of ZMPC have been reported in the literature, for example, treatment of diabetes in \cite{grosman_zone_2010, gonzalez_stable_2020}, control of heat system inside a building in \cite{privara_model_2011}, control of coal-fired boiler-turbine generating systems in \cite{zhang_zone_2020}, and control of irrigation systems in \cite{mao_soil_2018} and \cite{huang_model_nodate}. 
	
	Closed-loop stability of the system is always an essential aspect to be considered in controller design, as unstable controllers can lead to drastic safety risks in process operation. The closed-loop stability theories of MPC and EMPC are well-developed and widely used in literature based on the Lyapunov stability theory \cite{angeli_average_2012, heidarinejad_economic_2012-1, liu_economic_2016, liu_lyapunov-based_2017, makhamreh_lyapunov-based_2019}. 
	However, most of these stability theories rely on the assumption of the existence of an optimal operating point for both MPC and EMPC. On the other hand, ZMPC is often considered to be a relaxation to conventional MPC with not many investigations done in terms of the stability property. The stability of ZMPC with a secondary economic objective is investigated in \cite{liu_economic_2018}. An extension of \cite{liu_economic_2018} considered the presence of disturbance, of which a ZMPC that tracks an alternative economic zone is proposed \cite{decardi-nelson_robust_2021}. In both works, the stability theories rely on the assumption that there exists an optimal steady-state operating condition. On the other hand, a generalized ZMPC was proposed with no assumption on steady state optimal operating point in \cite{liu_model-predictive_2019}, however without considering any system disturbance or noise. 
	
	In this work, we propose a robust ZMPC with guaranteed convergence in the presence of bounded disturbance, which is an extension of the results obtained in \cite{liu_model-predictive_2019}. The proposed design rejects the effect of disturbance by modifying the target set. The modified target set is a subset of the actual target set, which the shrinkage is proved to be upper-bounded. A terminal constraint is employed to ensure the closed-loop stability, which forces the system state to converge to a forward control invariant set (CIS) inside the modified target set. Based on the stability theory, a practical guideline for determining the modified target zone is proposed.
	
	On top of the generalized robust ZMPC, we consider a secondary economic objective. Instead of introducing economic zones as proposed in \cite{decardi-nelson_robust_2021}, we prioritize the zone tracking objective and establish stability based on the ZMPC alone. This implies that the economic objective is optimized only if the zone-tracking objective is satisfied, which provides a larger feasible region and more flexibility, as the economic objective can be modified without affecting the stability property of the controller. 
	
	The performance of the proposed ZMPC is validated through simulations based on a continuous stirred tank reactor (CSTR). The performance of the generalized ZMPC proposed in \cite{liu_model-predictive_2019} is employed as the benchmark reference. To show the effectiveness of the proposed design, simulations based on alternative controller formulations are performed and the results are compared with those obtained based on the proposed design. The effectiveness of the proposed practical guideline is tested through simulations with different values of the tuning parameter. A remark regarding the selection of the tuning parameter is summarized.
	
	The remainder of this script is organized as follows: 
	Section \ref{section:Prelim} introduces the preliminaries and the system of interest, Section \ref{section:Proposed} presents the proposed ZMPC formulation, of which the stability proof is presented in Section \ref{section:stability}. Based on the stability theory, an algorithm for estimating the modified target set is designed in Section \ref{section:practical}. Section \ref{section:econ} introduces the proposed ZMPC with additional economic objective. The simulation results obtained based on the CSTR are presented in \ref{section:sim}. Finally, concluding remarks and ideas for future works are summarized in Section \ref{section:conclusion}.
	
	\section{Preliminaries}
	\label{section:Prelim}
	\subsection{Notation}
	Throughout this work, $|\cdot|$ denotes the Euclidean norm of a scalar or a vector. 
	The operator $|| \cdot ||_n$ denotes the $n$-norm of a scalar or a vector.
	$\m I_M^N$ represents the set of integers from $M$ to $N$ $\{M, M+1, \cdots, N\}$.
	$\m I_{\geq 0} = \{0,1,2,...\}$ denotes the set of non-negative integers.
	For two sets $A$ and $B$, the set subtraction is defined as $A \backslash B = \{a \in A|a \notin B\}$.
	The operators $\max {\{\cdot\}}$ and $\min {\{\cdot\}}$ find the maximum and minimum of each element in a vector variable respectively.
	
	The operator $a \odot b$ represents element-wise multiplication between vectors $a$ and $b$.
	\subsection{Description and problem formulation} 
	The following discrete-time nonlinear system is considered in this work:
	\begin{equation}
		\label{act_sys}
		x(n+1) = f(x(n), u(n), w(n))
	\end{equation}
	where $x(\cdot) \in \m X \subseteq \m R^{n_x}$ denotes the state vector, $u(\cdot) \in \m U \subseteq \m R^{n_u}$ denotes the control input vector, and $w(\cdot) \in \m W \subseteq \m R^{n_w}$ denotes the disturbance vector. $n \in \m I_{\geq 0}$ denotes the time step. 
	
	A few assumptions are enforced on the system of interest throughout this work and are listed below:
	\begin{assumption}
		\label{constraint_compact}
		The constraint sets $\m X$ and $\m U$ are compact and coupled such that:
		\[(x(n), u(n)) \in \m Z \subseteq \m X \times \m U\] 
		Furthermore, the sets can be expressed as following element-wise constraints:
		\[\m X :=\left\{x\Big{|} x_{lb} \leq x \leq x_{ub}\right\}, \: \m U :=\left\{u\Big{|} u_{lb} \leq u \leq u_{ub}\right\}  \]
		where $x_{lb}$, $x_{ub}$, $u_{lb}$, and $u_{ub}$ are constant vectors, with each elements being the element-wise lower and upper bound of the state and input. The expressions imply that the constraint sets are polyhedrons in the corresponding space bounded by hyperplanes.
	\end{assumption}
	\begin{assumption}
		\label{bounded_distb}
		The disturbance vector $w$ is bounded and contains the origin in its domain:
		\begin{equation}
			\m W := \{w \in \m R^{n_w}: ||w||_\infty \leq \theta, \theta > 0 \}
		\end{equation}
		such that the infinity norm of $w$ is bounded by a positive constant $\theta$.
	\end{assumption}
	\begin{assumption} 
		\label{continuous}
		The function $f: \m R^{n_x} \times \m R^{n_u} \times \m R^{n_w} \rightarrow \m R^{n_x}$ is locally Lipschitz with respect to $x$ and $w$ for all $x \in \m X$, $u \in \m U, w \in \m W$. This implies there exist positive constants $L_x$ and $L_w$ such that:
		\begin{equation}
			|f(x,u,w) - f(z,u,0)|  \leq L_w|w| + L_x|x -z|
		\end{equation}
		which indicates that the function $f$ cannot change drastically as $x$ and $w$ changes.
	\end{assumption}
	The control objective of this work is to drive the state of system (\ref{act_sys}) into a tracking target set $\m X_t$ and keeps it inside $\m X_t$ thereafter in the presence of process uncertainties. The target set $\m X_t$ can be expressed as the following element-wise constraints:
	\begin{equation}
		\m X_t :=\left\{x\Big{|} x^t_{lb} \leq x \leq x^t_{ub}\right\} 
	\end{equation}
	where $x^t_{lb}$ and $x^t_{ub}$ are constant vectors, with each element being the element-wise lower and upper bound on the state.
	
	To achieve the control objective, we propose a robust ZMPC with guaranteed closed-loop stability in the Lyapunov sense, which is introduced in Section \ref{section:Proposed}. The proposed robust ZMPC is an extension of the generalized ZMPC proposed in \cite{liu_model-predictive_2019}, which is introduced first in Section \ref{section:GenZMPC} for the completeness. 
	
	\subsection{A generalized ZMPC formulation}
	\label{section:GenZMPC}
	The generalized ZMPC proposed in \cite{liu_model-predictive_2019} is presented in this section. This generalized ZMPC assumes perfect knowledge regarding the system dynamics (i.e. $x(n+1) = f(x(n), u(n), 0)$). Throughout this work, we refer to this generalized ZMPC as the nominal ZMPC, which is used as the referencing benchmark for the performance validation of our proposed ZMPC. 
	The generalized ZMPC formulation for time instant $n$ is defined as follows:
	\begin{subequations}\label{eqn:ZMPC}
		\begin{align}
			V_N^0(x(n)) = &\min\limits_{u(0), \cdots, u(N-1) }    \sum_{i=0}^{N-1} \ell_z(\hat{x}(i)) \label{eqn:ZMPCa}  \vspace{2mm} \\ 
			{\rm s.t.~} &  \hat{x}(i+1) = f(\hat{x}(i),u(i), 0),~~ i \in \m I_{0}^{N-1} \vspace{2mm} \label{eqn:ZMPCb}   \\
			& \hat{x}(0)=x(n) \label{eqn:ZMPCc} \vspace{2mm}\\
			& \hat{x}(i) \in \m X, ~~i \in \m I_{0}^{N-1}\label{eqn:ZMPCd} \vspace{2mm}\\
			& u(i) \in \m U, ~~i \in \m I_{0}^{N-1}\label{eqn:ZMPCe} \\
			& \hat{x}(N) \in  \m X_f \label{eqn:ZMPCf}
		\end{align}
	\end{subequations}
	where (\ref{eqn:ZMPCb}) is the nominal process model constraint, (\ref{eqn:ZMPCc}) is the initial condition constraint, and (\ref{eqn:ZMPCd}) and (\ref{eqn:ZMPCe}) are the state and input constraints, respectively. (\ref{eqn:ZMPCf}) denotes the terminal constraint, where $\m X_f$ denotes the terminal set. $N$ is a positive constant that represents the control horizon of the controller. $\hat{x}(i)$ denotes the predicted states over the prediction horizon for $i \in \m I_{0}^{N-1}$. $\ell_z (\cdot)$ represents the zone-tracking objective and is defined as follows:
	\begin{subequations} \label{eqn:l_z}
		\begin{align}
			\ell_z(z) = \min\limits_{z_z} \: & c_1 ||z - z_z||_1 + c_2 ||z - z_z||_2^2 \\
			\label{eqn:l_z_2}
			s.t. \: & z_z \in \m X_t
		\end{align}
	\end{subequations}
	where $c_1$ and $c_2$ are non-negative weighting factors, $\m X_t \subset \m X$ denotes the target zone, and $z_z$ is a slack variable that is forced to stay inside the target zone by (\ref{eqn:l_z_2}). The zone-tracking objective is a weighted summation of the 1-norm and squared 2-norm of the minimum difference between the actual state and the slack variable, which is a representation of the distance between the system state $x$ and the target set $\m X_t$. When the system state is outside the target set, $\ell_z$ is positive. When the system state converges to the target set, $\ell_z$ equals to zero.
	
	Taking the definition of (\ref{eqn:l_z}) into consideration, the objective of the controller (\ref{eqn:ZMPC}) is to find the optimal input sequence for $N$ future steps such that the system state is driven towards and kept inside the target zone while constraints (\ref{eqn:ZMPCb}) - (\ref{eqn:ZMPCe}) are satisfied. In addition, the terminal state $x_N$ is forced to converge to $\m X_f \subseteq \m X_t^M \subseteq \m X_t$, where $\m X_t^M$ denotes the largest forward control invariant set (CIS) inside $\m X_t$. The definition of a forward CIS is provided as follows:  
	\begin{definition}	 (Forward control invariant set \cite{blanchini_set_1999})
		A set $\m X_r \subseteq \m X$ is said to be a forward control invariant set (CIS) for the system $x(n+1) = f(x(n), u(n), 0)$ if there exists a control policy $u(n) = \mu (x(n))$ for every $x(n) \in \m X_r$ such that $x(n+1) = f(x(n), \mu(x(n)), 0) \in \m X_r$. 
	\end{definition}
	We refer to the forward CIS as CIS hereafter for simplicity. The nominal ZMPC is proved to be closed-loop stable in the Laypunov sense when the system model is exactly known. In this work, we extend the design of the nominal ZMPC to consider nonlinear systems with model-plant mismatch in the form of process disturbance as shown in (\ref{act_sys}). A robust ZMPC with guaranteed closed-loop stability is proposed in the following section.
	\section{The proposed robust ZMPC}
	\label{section:Proposed}
	
	The proposed robust ZMPC is presented in this section. The effect of process uncertainty is encountered through modifying the tracking target set and the terminal set. 
	The proposed robust ZMPC is designed based on the nominal system $\tilde x(n+1) = f(x(n), u(n), 0)$, as the value of the disturbance $w(n)$ is unknown at any given time $n \in \m I_{\geq 0}$. 
	The proposed ZMPC design for time instant $n$ is presented as follows:
	\begin{subequations}\label{eqn:ZMPC2}
		\begin{align}
			V_N^0(x(n)) = & \min\limits_{u(0), \cdots, u(N-1) }   \sum_{i=0}^{N-1} \tilde \ell_z(\tilde x(i)) \label{eqn:ZMPC2a}  \vspace{2mm} \\ 
			{\rm s.t.~} &  \tilde x(i+1) = f(\tilde x(i),u(i), 0),~~ i \in \m I_{0}^{N-1} \vspace{2mm} \label{eqn:ZMPC2b}   \\
			& \tilde x(0)=x(n) \label{eqn:ZMPC2c} \vspace{2mm}\\
			& \tilde x(i) \in \m X, ~~i \in \m I_{0}^{N-1}\label{eqn:ZMPC2d} \vspace{2mm}\\
			& u(i) \in \m U, ~~i \in \m I_{0}^{N-1}\label{eqn:ZMPC2e} \vspace{2mm}\\
			& \tilde x(N) \in \tilde{\m X}_f \label{eqn:ZMPC2f}
		\end{align}
	\end{subequations}
	where $\tilde \ell_z(\cdot)$ is defined as follows:
	\begin{subequations} \label{eqn:l_z2}
		\begin{align}
			\tilde \ell_z(z) = \min\limits_{z_z} \: & c_1 ||z - z_z||_1 + c_2 ||z - z_z||_2^2 \label{eqn:l_z2a} \vspace{2mm}\\
			s.t. \: & z_z \in \tilde{\m X}_t \label{eqn:l_z2b}
		\end{align}
	\end{subequations}
	Similar notations employed in (\ref{eqn:ZMPC}) and (\ref{eqn:l_z}) are adopted in the proposed controller (\ref{eqn:ZMPC2}) and (\ref{eqn:l_z2}) except that $\tilde x$ is used to represent the nominal state predicted by the nominal model. Compare to the nominal ZMPC, the key changes took place in the proposed design are the modified target set $\tilde{\m X}_t$ and the terminal set $\tilde{\m X}_f$ in (\ref{eqn:l_z2b}) and (\ref{eqn:ZMPC2f}), respectively. The modified target set is a subset of the original target set ($\tilde{\m X}_t \subset \m X_t$). The terminal constraint is needed to ensure the closed-loop stability with a CIS inside the modified target zone $\tilde{\m X}_t$. 
	
	The proposed ZMPC is applied to system (\ref{act_sys}) in the receding horizon fashion. At each time instant $n \in \m I_{\leq0}$, the optimal input $u(i|n)^*, ~i\in \m I_{0}^{N-1}$ for $N$ future steps is obtained by solving the optimization problem (\ref{eqn:ZMPC2}). Only the optimal input for $i=0$ is applied to the system at time $n$. At time $n+1$, the optimization problem (\ref{eqn:ZMPC2}) is evaluated again with updated initial condition.
	\begin{figure}[!t]
		\centering
		\includegraphics[width=0.5\columnwidth]{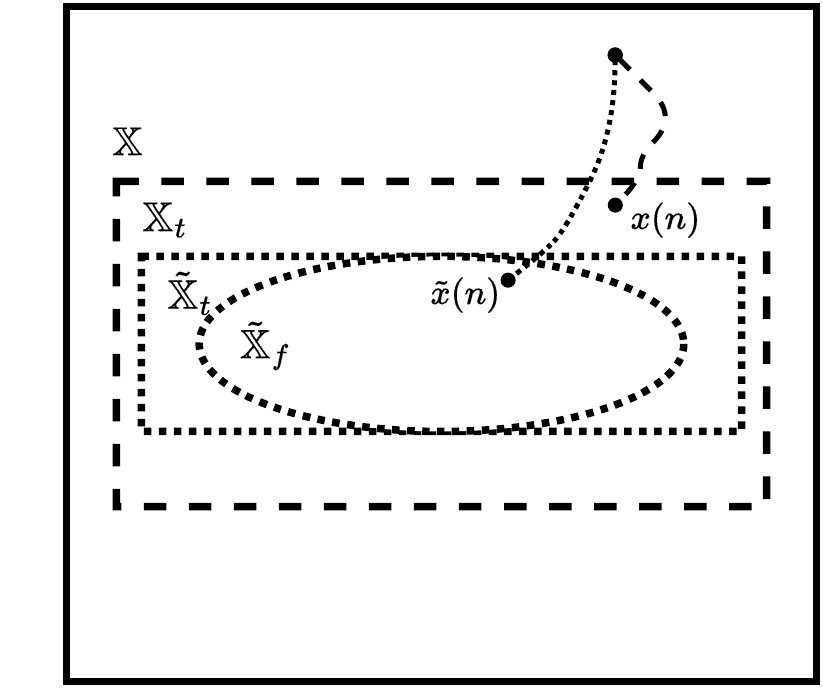}
		\caption{Potential nominal state trajectory (dotted line) and actual state trajectory (dashed line) of the closed-loop system under the proposed ZMPC.}
		\label{fig:compare}
	\end{figure}

	Figure \ref{fig:compare} presents potential trajectories of the closed-loop nominal state and the actual state based on the proposed robust ZMPC. The solid box represents the state space $\m X$, while the rectangles bounded by the dashed and the dotted lines represent the actual target set $\m X_t$ and the modified target set $\tilde{\m X}_t$, respectively. The ellipsoid bounded by the dotted line denotes the largest CIS $\tilde{\m X}_t^M$ inside the modified target set, which is used as the terminal set $\tilde{\m X}_f$ of the proposed controller. Oftentimes, a nominal ZMPC tends to drive the system state to the boundary of the target set, which likely leads to target set violation in the presence of process uncertainties. By ensuring the closed-loop convergence of the nominal state to the modified target set, which is a subset of the actual target set, a buffer zone is provided to reject the effect of process uncertainties. The terminal constraint (\ref{eqn:ZMPC2f}) ensures that the nominal state will stay inside the control invariant terminal set $\tilde{\m X}_f$ and thus $\tilde{\m X}_t$ once it has first converged. More details regarding the closed-loop stability of the proposed controller are discussed in the following section.  
	
	\begin{remark}
		Throughout the manuscript, we use $x(\cdot)$ to denote the true system state at a given time step. Contrarily, $\tilde x(\cdot)$ denotes the system state predicted based on the knowledge available to the controller. 
		Following similar ideas, the set $\m X_t$ reflects the true zone control objectives we have on the system of interest, which is referred to as the actual target set. For example, the objective is to keep the reactor temperature in a certain range. The set $\tilde{\m X}_t$ in comparison, is mathematically determined based on the actual target set with no physical significance and is referred to as the modified target set. 
		Furthermore, we define the set $\m X_t^M$ and $\tilde{\m X}_t^M$ as the largest CIS inside $\m X_t$ and $\tilde{\m X}_t$, respectively.
	\end{remark}

	

	
	\section{Stability analysis}
	\label{section:stability}
	In this section, we investigate the closed-loop stability of the proposed robust ZMPC (\ref{eqn:ZMPC2}). First, we introduce the definition of the $N$-step reachable set, which is essential in characterizing the feasibility of the proposed robust ZMPC.

	\begin{definition} ($N$-step reachable set \cite{liu_model-predictive_2019}) 
		We use $\m X_N(\m X, \m X_f)$ to denote the set of states in $\m X$ that can be steered to $\m X_f$ in $N$ steps while satisfying the state and input constraints $(x,u) \in \m Z$. That is,
		\[
		\m X_N(\m X, \m X_f) =\big\{ x(0) \in \m X \; | \; \exists\:(x(n),u(n)) \in \m Z,  n \in  \m I_0^{N-1}, x(N) \in \m X_f   \big\}
		\]
	\end{definition} 
	
	The following assumption is employed to ensure the feasibility of the proposed robust ZMPC.
	\begin{assumption}
		\label{reachable}
		The terminal set $\tilde{\m X}_f$ and the $N$-step reachable set $\m X_N(\m X, \tilde{\m X}_f)$ are compact and nonempty. 
	\end{assumption}
	
	For the completion of this work, here we summarize the essential results obtained in \cite{liu_model-predictive_2019} regarding the nominal ZMPC as Theorem \ref{thrm:Su}.
	\begin{theorem} (Stability of the nominal ZMPC \cite{liu_model-predictive_2019})
		\label{thrm:Su}
		If Assumptions \ref{constraint_compact} and \ref{reachable} hold, $x_0 \in \m X_N(\m X, \m X_f)$, $w(n) = 0$, and in addition, the optimal value function $V_N^0(x(n))$ is locally continuous on $\m X_t$, then:
		(\rNum{1}) the zone MPC is recursively feasible with $x(n) \rightarrow \m X_t^M$,
		(\rNum{2}) the terminal set $\m X_t^M$ is asymptotically stable, (\rNum{3}) the following inequality holds:
		\begin{equation}
			\label{eqn:nominal_V_decrease}
			V_N^0(\tilde x(n+1)) - V_N^0(\tilde x(n)) \leq -\ell_z(\tilde x(n)), \;\; n \in \m  I_{\geq 0}
		\end{equation}
		where $V_N^0$ is the optimum control objective function and is a Lyapunov function with respect to the set $\m X_t^M$, and $\ell_z(\cdot)$ denotes the zone-tracking objective function defined in (\ref{eqn:l_z}).
	\end{theorem}

	As the proposed ZMPC is applied to the system of interest in the receding horizon fashion, the optimization problem is solved at each time step with the updated information regarding the actual state at the initial time step over the prediction horizon (i.e. $ \tilde x(n) = x(n)$). This implies only the one-step-ahead deviation caused by the plant-model mismatch between the nominal model and the actual model needs to be considered. The following Proposition provides an upper bound on the one-step-ahead deviation of the predicted state from the actual state:
	
	\begin{proposition}(c.f. \cite{decardi-nelson_robust_2021})
		\label{prop1}
	Consider the actual system (\ref{act_sys}) and the nominal system with $w(n) = 0$.
	If Assumption \ref{bounded_distb} is satisfied, starting from a known initial condition $x(n)$, the deviation of the actual state $x(n+1)$ from the predicted state $\tilde x(n+1)$ in one time step is bounded:
	\begin{equation}
		\label{eqn:deviation}
		|x(n+1) - \tilde{x}(n+1)| \leq \sqrt{n_x} L_w \theta,~\forall x(n+1),~\tilde x(n+1) \in \m X
	\end{equation}
	\end{proposition}
	Based on the upper bound of the one-step-ahead deviation in the state, Proposition \ref{prop2} provides an upper bound on the progression of the Lyapunov function $V_N^0(\cdot)$ in one time step from $x(n)$ to $x(n+1)$.
	\begin{proposition}
		\label{prop2}
		Consider the Lyapunov function $V_N^0(x)$. There exist positive constants $K_V$ and $H$ such that: 
		\begin{equation}
			\label{eqn:act_V_decrease}
			V_N^0(x(n+1)) - V_N^0(x(n)) \leq -\tilde{\ell}_z(\tilde x(n)) + f_V(\sqrt{n_x} L_w\theta) 
		\end{equation}	
		where $n_x$ and $\theta$ are known parameters, $\tilde{\ell}_z(\cdot)$ denotes the zone-tracking objective function defined in (\ref{eqn:l_z2}), and $f_V$ is a function defined as follows:
		\begin{equation*}
			f_V(x) = K_V x  + H x^2
		\end{equation*}
		where $K_V$ and $H$ are positive constants.
	\end{proposition}
		{\bf Proof.}
		Upon applying Taylor expansion to $V_N^0(x)$ around the predicted state $\tilde x(n+1)$, the following relation can be obtained:
		\begin{equation}
			\label{eqn:TE}
			\begin{split}
				&V_N^0(x(n+1)) = V_N^0(\tilde x(n+1)) +  \\
				&\left.\frac{\partial V_N^0}{\partial x} \right\vert_{\tilde x(n+1)} |x(n+1)- \tilde x(n+1)| + H.O.T.
			\end{split}
		\end{equation}
		where $H.O.T.$ includes the higher order terms of the Taylor expansion. A positive constant $H$ can be found for $x \in \m X$ such that the following holds:
		\begin{equation}
			\label{eqn:HOT}
			H.O.T. \leq H |x(n+1)- \tilde x(n+1)|^2
		\end{equation}
		Consider (\ref{eqn:TE}) and (\ref{eqn:HOT}) correspondingly, we can obtain the following inequality regarding the difference in $V_N^0$ due to the effect of disturbance in one step:
		\begin{equation}
			\label{ineq:TE}
			\begin{split}
				V_N^0(x&(n+1))  \leq V_N^0(\tilde x(n+1)) \\
				& +  \left.\frac{\partial V_N^0}{\partial x} \right\vert_{\tilde x(n+1)} |x(n+1)- \tilde x(n+1)| \\
				&+  H |x(n+1)- \tilde x(n+1)|^2
			\end{split}
		\end{equation}
		If Assumptions \ref{continuous} and \ref{reachable} are satisfied, there exists a positive constant $K_V$ such that the magnitude of the partial derivative $\frac{\partial V_N^0}{\partial x}$ is bounded by it:
		\begin{equation}
			\label{ineq:HOT}
			\left\vert\frac{\partial V_N^0}{\partial x} \right\vert \leq K_V
		\end{equation}
		Define $f_V(x) = K_V x  + H x^2$. Based on (\ref{ineq:TE}) and (\ref{ineq:HOT}), applying Proposition \ref{prop1}, the following relationship can be derived:
		\begin{equation}
			\label{eqn:deviation_V}
			V_N^0(x(n+1)) \leq V_N^0(\tilde x(n+1)) + f_V(\sqrt{n_x} L_w\theta) 
		\end{equation}
		Taking into account of (\ref{eqn:nominal_V_decrease}) and (\ref{eqn:deviation_V}), plus the fact that $x(n) = \tilde x(n)$, the relationship in (\ref{eqn:act_V_decrease}) can be obtained and this proves Proposition \ref{prop2}. $\blacksquare$

	To derive the amount of shrinkage required in the target set that ensures the closed-loop convergence of the system in the presence of disturbance, we introduce the following definition of Lyapunov function level set:
	\begin{definition} (Level set of Lyapunov function)
		The set $\Omega_\rho$ is defined to be the level set of the Lyapunov function $V_N^0(\cdot)$:
		\begin{equation}
			\Omega_\rho := \{x\in \m X: V_N^0(x) \leq \rho\}
		\end{equation}
	\end{definition}
	Let $\Omega_{\rho_{max}}$ be the largest level set of $V_N^0$ inside $\m X_t^M$:
	\begin{equation}
		\rho_{max} := max\{\rho: x\in \m X_t^M, V_N^0(x) \leq \rho\}
	\end{equation}
	
	Define $\Omega_{\rho_{e}} \subset \Omega_{\rho_{max}}$ that satisfies:
	\begin{equation}
		\label{level_set_shrink}
		\rho_e = \rho_{max} - f_V(\sqrt{n_x} L_w\theta) 
	\end{equation}
	\begin{assumption}
		\label{nonempty_O}
		 $\Omega_{\rho_{e}}$ is nonempty.
	\end{assumption}
	\begin{theorem}
		\label{thrm:1step}
		Consider system (\ref{act_sys}) under the control of the proposed ZMPC (\ref{eqn:ZMPC2}). 
		If $x(n) \in \Omega_{\rho_{e}}$ and Assumptions \ref{continuous} - \ref{nonempty_O} are satisfied, then it is guaranteed that $x(n+1) \in \Omega_{\rho_{max}}$. 
	\end{theorem}
	{\bf Proof.}
	At a given time instant $n$, with the initial condition $x(n) \in \Omega_{\rho_{e}}$, it is guaranteed that there exists a feasible control policy such that the nominal state at the next time step $\tilde x(n+1) \in \Omega_{\rho_{e}}$, as $\Omega_{\rho_{e}}$ is control invariant. Take the inequality expression (\ref{eqn:deviation_V}) into consideration, the deviation in the value of Lyapunov function $V^0_N$ from $\tilde x(n+1)$ to $x(n+1)$ is bounded by $f_V(\sqrt{n_x} L_w\theta)$. Recall that $\Omega_{\rho_{e}}$ is a shrunk Lyapunov function level set with respect to $\Omega_{\rho_{max}}$, with the shrinkage defined in the Lyapunov function value by $f_V(\sqrt{n_x} L_w\theta)$, which proves Theorem \ref{thrm:1step}. $\blacksquare$ 
	
	Theorem \ref{thrm:1step} indicates the convergence of the actual state into a control invariant subset of the actual target set $\m X_t$ if the initial state $x(n) \in \Omega_{\rho_{e}}$. The following theorem provides statements on the upper bound of the shrinkage amount in the target set required in the proposed ZMPC design such that the state is guaranteed to be driven into $\Omega_{\rho_{e}}$.

	\begin{definition}(The smallest polyhedron set around a known set)
		We define the smallest polyhedron set bounded by orthogonal hyperplanes $\m P(\m S)$ around a given set $\m S\subseteq \m R^{n_s}$ as follows:
		\begin{equation}
			\m P(\m S) = \{s|s_{lb} \leq s\leq s_{ub}\}
		\end{equation}
		where $s_{lb},s_{ub} \in \m R^{n_s}$ are vectors with each element $s_{lb}^{i},~i = 1,2,\cdots,{n_s}$ and $s_{ub}^{i},~i = 1,2,\cdots,{n_s}$ being the lower and upper bound of each dimension of the set $\m S$:
		\begin{subequations}\label{eqn:lbub_set}
			\begin{align}
				s_{lb}^{i} & = \min\{s^{i}|s\in \m S\},~i = 1,2,\cdots,{n_s} \\ 
				s_{ub}^{i} & = \max\{s^{i}|s\in \m S\},~i = 1,2,\cdots,{n_s}
			\end{align}
		\end{subequations}
		where $s^{i}$ is the $i$-th element of the vector $s$.
	\end{definition}
	\begin{theorem}
		Consider system (\ref{act_sys}) under the control of the proposed ZMPC (\ref{eqn:ZMPC2}). If Assumptions \ref{constraint_compact} - \ref{nonempty_O} are satisfied, $x(0) \in \m X_N(\m X, \tilde{\m X}_f)$, and the following relationship holds:
		\begin{equation}
			\label{eqn:convergence}
			-\tilde{\ell}_z(x) + f_V(\sqrt{n_x} L_w\theta) < 0,~\forall x\in\m X_N(\m X, \tilde{\m X}_f) \backslash \Omega_{\rho_{e}}
		\end{equation}
		then there exists $\overline{\varepsilon}, \underline{\varepsilon} \in \mathbb{R}^{n_x}$ with non-negative elements that define the modified target zone $\tilde{\m X}_t$ as follows:
		\begin{equation}
			\tilde{\m X}_t= \{x: \tilde x_{lb}^t \leq x \leq \tilde x_{ub}^t| \tilde x_{lb}^t - x_{lb}^t \leq \underline{\varepsilon}, x_{ub}^t - \tilde x_{ub}^t \leq \overline{\varepsilon} \}
		\end{equation} 
		such that the proposed ZMPC controller is recursively feasible and guarantees the closed-loop convergence of the system state into the actual target zone $\m X_t$. Note that $x_{lb}^t$ and $x_{ub}^t$ are known vectors containing the element-wise lower and upper bound of the actual target set $\m X_t$ as defined in Assumption \ref{constraint_compact}.
	\end{theorem}
	{\bf Proof.}
			%
	We define the modified target set as the smallest polyhedron set around $\Omega_{\rho_{e}}$ (i.e. $\tilde{\m X}_t = \m P(\Omega_{\rho_{e}})$) and the corresponding terminal set $\tilde{\m X}_f$ as the largest CIS inside $\tilde{\m X}_t$. If Assumption \ref{nonempty_O} holds, it means there exists a valid set of $\tilde x_{lb}^t$ and $\tilde x_{ub}^t$ such that $\tilde{\m X}_t \subset \m X_t$ is bounded and nonempty. Recall that the actual target set $\m X_t$ is bounded, meaning that the elements in vectors $\overline{\varepsilon}, \underline{\varepsilon} \in \mathbb{R}^{n_x}$ are bounded and non-negative.
	
	Take the relationship defined by (\ref{eqn:convergence}) into consideration. The value of the Lyapunov function $V_N^0$ is guaranteed to be decreasing until the state converges to $\Omega_{\rho_{e}}$, thus ensuring the convergence of the nominal state $\tilde x$ into $\Omega_{\rho_{e}}$.
	
	In the presence of disturbance, the actual state $x$ may deviate from $\Omega_{\rho_{e}}$ after the first convergence of the nominal state $\tilde x$. Taking Theorem \ref{thrm:1step} into consideration, it is guaranteed that with $x(n) \in \Omega_{\rho_{e}}$, $x(n+1) \in \Omega_{\rho_{max}} $, meaning the actual state is kept inside $\Omega_{\rho_{max}} \subset \m X_t^M \subset \m X_t$ in one step. 
	As $\Omega_{\rho_{e}} \subset \Omega_{\rho_{max}} \subset \m X_N(\m X, \tilde{\m X}_f)$, the actual state that has deviated in one step due to the effect of disturbance is guaranteed to be driven back to $\Omega_{\rho_{e}}$ in finite steps as (\ref{eqn:convergence}) holds. This indicates that the proposed ZMPC controller is able to drive the system state into the actual target set $\m X_t$ and keeps it inside.  $\blacksquare$

	\begin{remark}
		By definition, any CIS inside the modified target set $\tilde{\m X}_t$ can be used as the terminal set $\tilde{\m X}_f$. However, in order to gain a controller with a larger feasibility region and more degrees of freedom, we define $\tilde{\m X}_f = \tilde{\m X}_t^M$. The largest CIS inside the modified target set is approximated using the algorithm proposed in \cite{decardi-nelson_computing_2021}. The algorithm provides an inner approximation of the largest CIS in the form of a polyhedron bounded by hyperplanes.
	\end{remark}
	\section{An algorithm for estimating the modified target set $\tilde{\m X}_t$}
	\label{section:practical}
	The stability proof in the previous section ensures the existence of an upper bound of the shrinkage amount on the actual target set, however, it is very challenging to obtain explicit expressions of the Lyapunov function and numerical values of the parameters. In order to make the proposed controller more practical in applications, we propose an algorithm for estimating and tuning the modified target set. 
	
	The proposed algorithm is inspired by a study on the problem in the context of linear systems. Consider a linear system as follows:
	\begin{equation}
		\label{linear_sys}
		x(n+1) = Ax(n) + Bu(n) + Ew(n)
	\end{equation}
	where $A$, $B$, and $E$ are matrices of the appropriate dimensions. Let us assume that the linear system is controllable and there exists the following quadratic Lyapunov function for the system:
	\begin{equation}
		V^0_N(x) = x^TPx
	\end{equation}
	where $P$ is a symmetric positive definite matrix. 
	Recall the Lipschitz continuity Assumption \ref{continuous}. If we substitute the linear system model (\ref{linear_sys}) into the left hand side of the inequality (\ref{continuous}), the following expression can be obtained:
	\begin{equation}
		|f(x,u,w) - f(x,u,0)| = |Ew| \leq L_w|w|
	\end{equation}
	By the property of norm, the following expression holds:
	\begin{equation}
		|Ew| \leq |E||w|
	\end{equation}
	Thus, we may define $L_w = |E|$ such that Assumption \ref{continuous} holds. Recall the inequality relation (\ref{eqn:deviation}):
	\[|x(n+1) - \tilde{x}(n+1)| \leq \sqrt{n_x} L_w \theta\]
	where $n_x$ is the dimension of the state, $\theta$ is the upper bound of the one-norm of the disturbance and is assumed to be known. Thus for the linear system, the right-hand side of the inequality can be determined explicitly. By calculating the term $\sqrt{n_x} L_w \theta$, a conservative approximation of the norm of the one-step-ahead difference between the actual state $x(n+1)$ and the nominal state $\tilde{x}(n+1)$ in the presence of disturbance is obtained. For the nonlinear system of interest, the matrix $E$ is essentially the sensitivity matrix of the state with respect to the disturbance (i.e. $\frac{\partial x}{\partial w}$). The value of the sensitivity matrix is a function of the state, input, and disturbance and can be calculated explicitly with a given combination of the variables. 
	
	Practically, based on inequality (\ref{eqn:deviation}), we propose the following definition of the 
	shrinkage amount $s$ in the state boundaries of the target zone for the nonlinear system: 
	\begin{equation}
		\label{eqn:shrinkage}
		s = \gamma x^d_{max}
	\end{equation}
	where $\gamma \in \m R^+$ is a risk factor that helps to tune how conservative the proposed controller is, and $x^d_{max}$ essentially estimates the largest potential effect the disturbance can have on the actual state compared to the nominal state in one-step-ahead prediction, of which the definition is inspired by (\ref{eqn:deviation}). $x^d_{max}$ can be calculated using Algorithm \ref{Alg_1} based on sensitivity analysis.
	\begin{algorithm}
		\caption{Estimation of the state deviation due to the presence of disturbance}
		\label{Alg_1}
		\begin{algorithmic}
			\Ensure $x \in \m X^M_t$, $u \in \m U$, $w \in \m W$
			\State $I_t \gets [\:]$
			\For{$i = 0:n_x-1$}
				\If {$x_{lb}<x^t_{lb}$ or $x_{ub}>x^t_{ub}$}
				\State $I_t \gets [I_t, 1]$
				\Else{ $I_t \gets [I_t, 0]$}
				\EndIf
			\EndFor
			\State Define $z \in \m R^{n_z} \gets [x, u]$, $n_z = n_x + n_u$ 
			\State Determine the sensitivity function $\frac{\partial x}{\partial w}: \m R^{n_z + n_w} \rightarrow \m R^{n_x \times n_w}$
			\State Find and store the maximum and minimum of each variables 
			\[z_{max}\in \m R^{n_z}\gets\max{\{z\}}, \: z_{min}\in \m R^{n_z} \gets \min{\{z\}}\]
			\State\[w_{max}\in \m R^{n_w}\gets\max{\{w\}}, \: w_{min}\in \m R^{n_w} \gets \min{\{w\}}\]
			\State Calculate the set of the accumulated sensitivity $\m X^d$ for the state variables with zone-tracking objective at each extreme point of $z$ and $w$:
			\[\m X^d := \left\{\left[\frac{\partial x}{\partial w}\Big{|}_{z^i, w^j} \cdot w\right] \odot I_t\; \Big{|} z^i \in [z_{max}^i, z_{min}^i], i \in \m I_0^{n_z-1}, w^j \in [w_{max}^j, w_{min}^j], j \in \m I_0^{n_w-1}\right\} \]
			\State Find the element $x^d_{max}$ with the largest norm in $\m X^d$ 
			\[x^d_{max} \gets \left\{x^d_{max} \;\Big{|} x^d_{max}\in \m X^d, |x^d_{max}| \geq |x^d|, \forall x^d \in \m X^d \right\}\] \\
			\Return $x^d_{max}$
		\end{algorithmic}
	\end{algorithm}

	In Algorithm \ref{Alg_1}, the set $\m X^d$ contains $2^{n_z + n_w}$ number of vectors. Each vector represents the amount of accumulated effects of the disturbance $w$ have on the state $x$ at one of the extremes of the joint space $\m X^M_t \times \m U \times \m W$. It is worth pointing out that the sensitivity matrix is computed for states inside the maximum CIS $\m X^M_t$, that is, inside the actual target set. This is because it has been proved in \cite{liu_model-predictive_2019} that for the nominal system, guaranteed convergence into the actual target set $\m X_t$ is equivalent to the convergence into its maximum control invariant subset $\m X^M_t$. Since the purpose of modifying the target set is to avoid zone violation once it has converged, we focus on rejecting the effect of disturbance when $x \in \m X^M_t$. Furthermore, the effect of disturbance is only investigated for the individual state variables that have a zone-tracking objective on them, as it is common that the zone-tracking objective is not enforced on all state variables. 

	The modified target zone can be calculated based on the shrinkage amount $s$ as follows:
	\begin{equation}
		\label{x_t_mod}
		\tilde{\m X}_t :=\left\{x\Big{|} x^t_{lb} + s \leq x \leq x^t_{ub} - s \right\} 
	\end{equation}

	
	\begin{remark}
		The inequality relationship (\ref{eqn:deviation}) is defined in terms of the norm of vectors and matrix. However, in practice, the shrinkage amount $s$ should be a vector that has the same dimension of the state, as the shrinkage required for each state needs to be determined individually. Thus, $x^d_{max}$ is selected to be the accumulated effect of $w$ on $x$ that has the largest norm, which mimics the idea of (\ref{eqn:deviation}), however in the state space. 
	\end{remark}

	\begin{remark}
		In Algorithm \ref{Alg_1}, the value of the sensitivity matrix needs to be calculated $2^{n_z+n_w}$ times, as every combination of the maximum and minimum of the state, input, and disturbance are considered. For higher-dimension systems, the computational effort may be significant. One way to reduce the computational cost is to utilize available physical knowledge in the implementation. For example, the optimal operating condition may be estimated based on the nominal model, or it is known to be on one edge of the zone. In these cases, the sensitivity matrix may be calculated only at the estimated optimal operating point, or at the extremes on one edge.
		
		Furthermore, (\ref{x_t_mod}) assumes that the target set shrinks in the same amount on the upper and lower bound. If the optimal operating point is known to be on one edge of the target zone, then the shrinkage can be applied to that edge only. This again helps to preserve a larger modified target zone and terminal zone, which helps to reduce the economic performance loss and feasibility issue. 
	\end{remark}

	
	\section{Consideration of economic objectives}
	\label{section:econ}
	A natural extension to the proposed ZMPC (\ref{eqn:ZMPC2}) is to include an economic objective, as the proposed design provides the system with more degrees of freedom to optimize additional objectives. The proposed ZMPC design that handles additional economic objective is presented as follows:
	\begin{equation} \label{eqn:ZMPC2_e}
		\begin{split}
			\min\limits_{u_0, \cdots, u_{N-1} } &   \sum_{i=0}^{N-1} \tilde \ell_z(\tilde x_i) + \ell_e(\tilde x_i, u_i) \vspace{2mm} \\
			{\rm s.t.~} &   \text{(\ref{eqn:ZMPC2b})}-\text{(\ref{eqn:ZMPC2f})}
		\end{split}
	\end{equation}
	
	With weighting factors being large enough on the zone control objective, namely $c_1$ in (\ref{eqn:l_z2}), the controller will prioritize the zone-tracking objective over the economic objective such that the economic objective will be optimized only after the system enters the target zone \cite{liu_model-predictive_2019}. Thus the stability properties of the controller (\ref{eqn:ZMPC2_e}) remain unchanged from that of (\ref{eqn:ZMPC2}). In the following section, simulation results based on (\ref{eqn:ZMPC2_e}) are presented. 
	
	\begin{remark}
		By tuning the parameter $\gamma$ in (\ref{eqn:shrinkage}), a trade-off between the size of the initial feasible region, the economic performance, and zone-tracking performance can be achieved due to changing target zone shrinkage $s$. A larger $\gamma$ leads to a smaller modified target zone and correspondingly, a smaller terminal zone. This enhances convergence to the actual target zone while potentially reducing the flexibility and feasibility of the controller, as the system is forced to converge into a smaller region at the end of the control horizon. Higher sacrifice in the economic performance may be observed as well. On the other hand, smaller $\gamma$ poses less effect on the controller feasibility and economic performance but the closed-loop system has a greater chance to violate the actual target zone. 
	\end{remark}
	\section{Simulations}
	\label{section:sim}
	In this section, we use a continuous stirred tank reactor (CSTR) as the benchmark process to demonstrate the effectiveness of the proposed design. After introducing the CSTR process (Section \ref{CSTR_process}) and the detailed setting of the proposed controller (Section \ref{controller_imp}), the performance of the proposed controller is first validated without and with a secondary economic objective with respect to the nominal ZMPC in Section \ref{validation}. The importance of shrinking the target set in the proposed controller design is verified in Section \ref{target_set}. The significance of the terminal constraint in ensuring stability is investigated in Section \ref{stab_validation}. Finally, the impact of the tuning parameter $\gamma$ is investigated in Section \ref{gamma}. 
	\subsection{Process description}
	\label{CSTR_process}
	The benchmark CSTR process is introduced in this section. An irreversible first-order exothermic reaction $A \rightarrow B$ takes place inside the CSTR reactor, where A and B are the reactants and the desired product, respectively. A cooling jacket is used to control the temperature of the CSTR. Assuming perfect mixing and constant mixture volume inside the reactor, the system can be described by the following ordinary differential equations:
	\begin{subequations}
		\begin{align}
			&\frac{dC_A}{dt} = \frac{q}{V} (C_{A_f} - C_A) - k_0 \exp{\Big(-\frac{E}{RT}\Big)}C_A \\
				&\frac{dT}{dt} = \frac{q}{V} (T_f - T) + \frac{UA}{V\rho C_p}(T_c - T)
				 + \frac{-\triangle H}{\rho C_p}k_0  \exp{\Big(-\frac{E}{RT}\Big)}C_A 
		\end{align}
	\end{subequations}
	where $C_A$ $[mol/L]$ denotes the molar concentration of the reactant, $T$ $[K]$ represents the temperature inside the reactor, $T_c$ $[K]$ represents the temperature of the cooling stream, and $C_{A_f}$ $[mol/L]$ and $T_f$ $[K]$ denote the molar concentration of the reactant and the temperature of the feed stream, respectively. $q$ $[L/min]$ represents the volumetric flow rate of the streams entering and leaving the reactor, $V$ $[L]$ and $\rho$ $[g/L]$ denote the volume and the density of the mixture inside the reactor respectively, and $k_0$, $E$ and $R$ are reaction-related parameters, namely the pre-exponential factor of the reaction rate, the activation energy, and the universal gas constant. $C_p$ denotes the specific heat capacity of the mixture, and $\triangle H$ and $UA$ denote the heat of reaction and the heat transfer coefficient between the reactor and the cooling jacket, respectively. The values of the parameters are adopted from \cite{decardi-nelson_robust_2021}.
	
	Recall the system defined in (\ref{act_sys}). For the CSTR example specifically, we define the system state vector to be $x = [C_A, T]^T$ and the control input to be $u = T_c$. The disturbance vector is defined as $w = [C_{A_f} - \bar{C}_{A_f}, T_f - \bar{T}_f]^T$, where $\bar{C}_{A_f} = 1.0 \, mol/L$ and $\bar{T}_f = 350\, K$ are the nominal values of the corresponding variables. 
	The following bounds are employed for the state, input, and disturbance, respectively:
	\begin{align}
			[0.0, 345.0]^T &\leq x \leq [1.0, 355.0]^T  \\
			285.0 &\leq u \leq 315.0 \\
			\label{eqn:w}
			[-0.1, -2.0]^T &\leq w \leq [0.1, 2.0]^T
	\end{align}
	\subsection{Controller setup and implementation}
	\label{controller_imp}
	The proposed ZMPC controller is implemented on the CSTR system without and with a secondary economic objective. Following the design proposed in (\ref{eqn:ZMPC2}) and (\ref{eqn:ZMPC2_e}), we define the zone control objective and economic objective. The zone control objective is to hold the reactor temperature $T$ between $348.0 \, K$ to $352.0\, K$. Recall the definition of $\ell_z$ in (\ref{eqn:l_z}), which implies the following:
	\[\m X_t = \{x: [0.0, 348.0]^T \leq x \leq [1.0, 352.0]^T\}\]
	The largest CIS inside the original and modified target zones are approximated using the algorithm proposed in \cite{decardi-nelson_computing_2021}.
	
	We computed the value of $x_{max}^d$ to be 0.511 for the states inside the largest CIS with the input and disturbances in their stated boundaries. It was found that the disturbance has the strongest effect on the state at the following point:
	\[x = [0.754, 352.0], \: u = 315.0, \: w = [0.1, 2]\]
	Unless otherwise mentioned, $\gamma = 1$ is used, which means $s = x^d_{max} = 0.511$.	
	The economic objective is to minimize the concentration of the reactant inside the reactor, i.e. $\ell_e = C_A$. For all the simulations, the control horizon $N=5$ unless otherwise mentioned.

	The value of the disturbance $w(n)$ at any time step is selected randomly inside the boundaries specified by (\ref{eqn:w}). Except for the CIS approximation that is carried out in Julia, all other simulations are performed in Python. The optimization problems are solved using the CasADi toolbox developed by \cite{Andersson2018}.
	\subsection{Control performance validation}
	\label{validation}
	\begin{figure}[t]
		\centering
		\includegraphics[width=0.8\columnwidth]{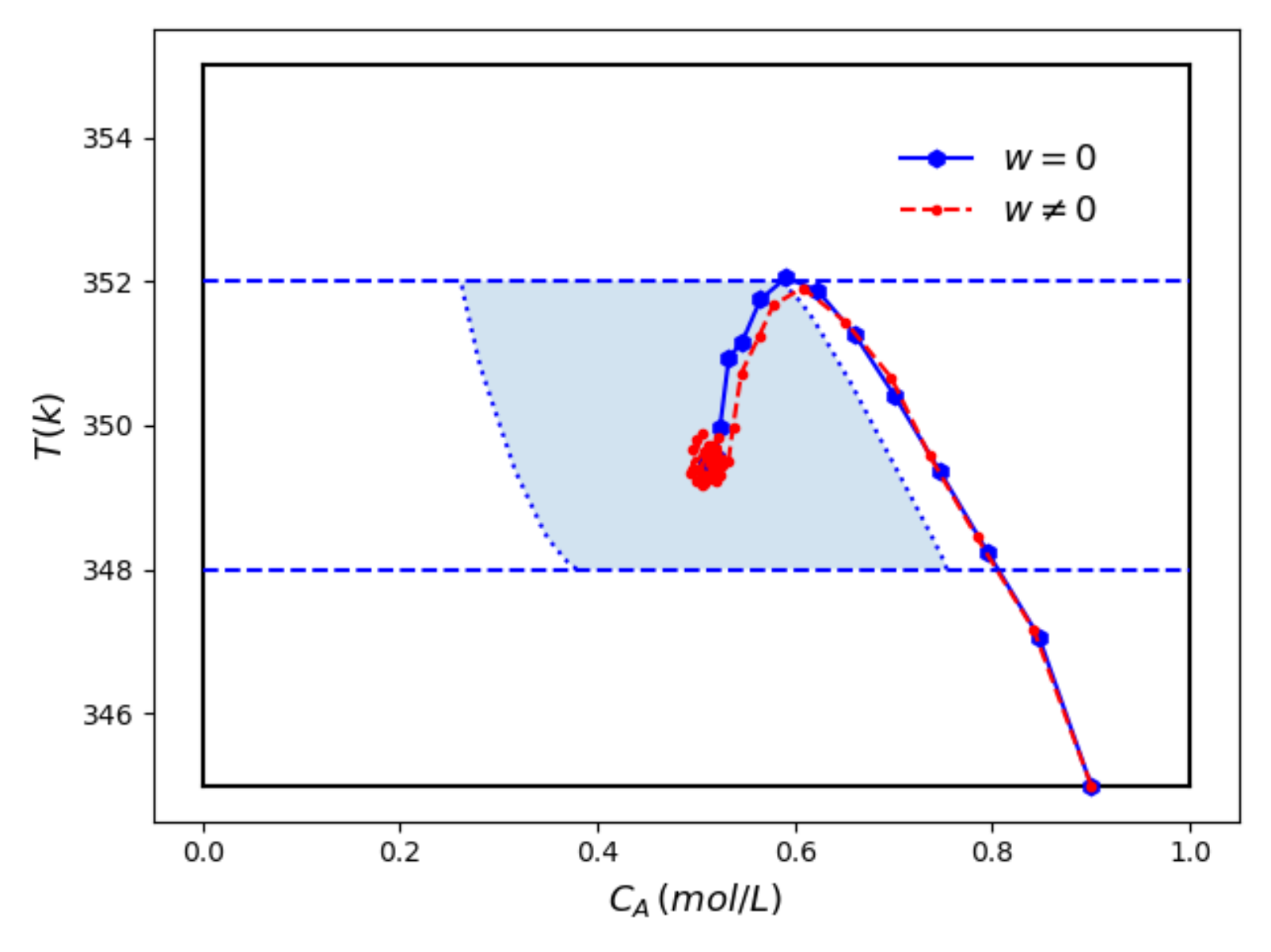}
		\caption{Proposed ZMPC without economic objectives.}
		\label{fig:le0}
	\end{figure}
	
	\begin{figure}[t]
		\centering
		\includegraphics[width=0.8\columnwidth]{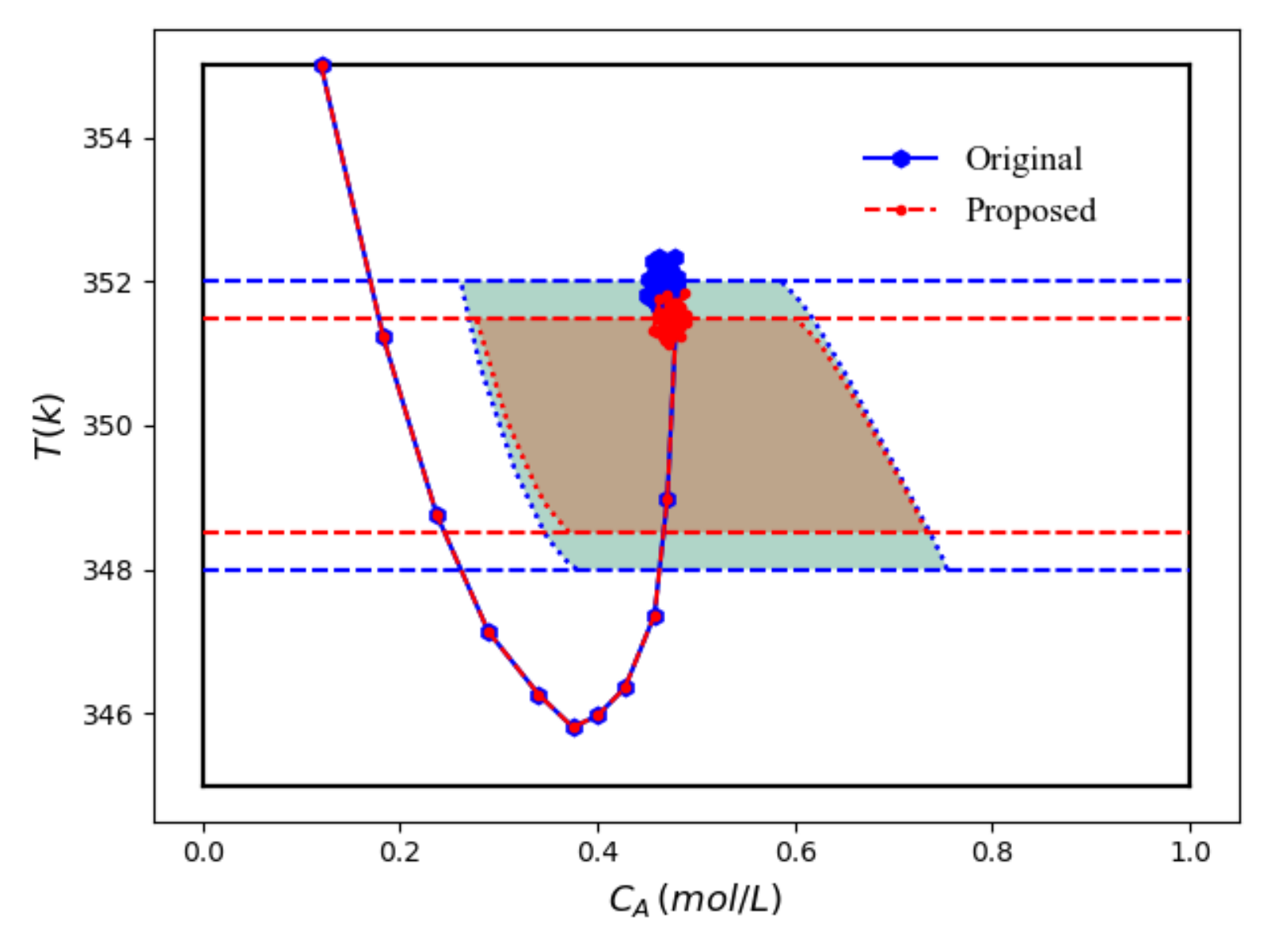}
		\caption{Proposed ZMPC v.s. the nominal ZMPC with economic objective, $x_0 = [0.12,355]$.}
		\label{fig:proposed1}
	\end{figure}
	
	\begin{figure}[t]
		\centering
		\includegraphics[width=0.8\columnwidth]{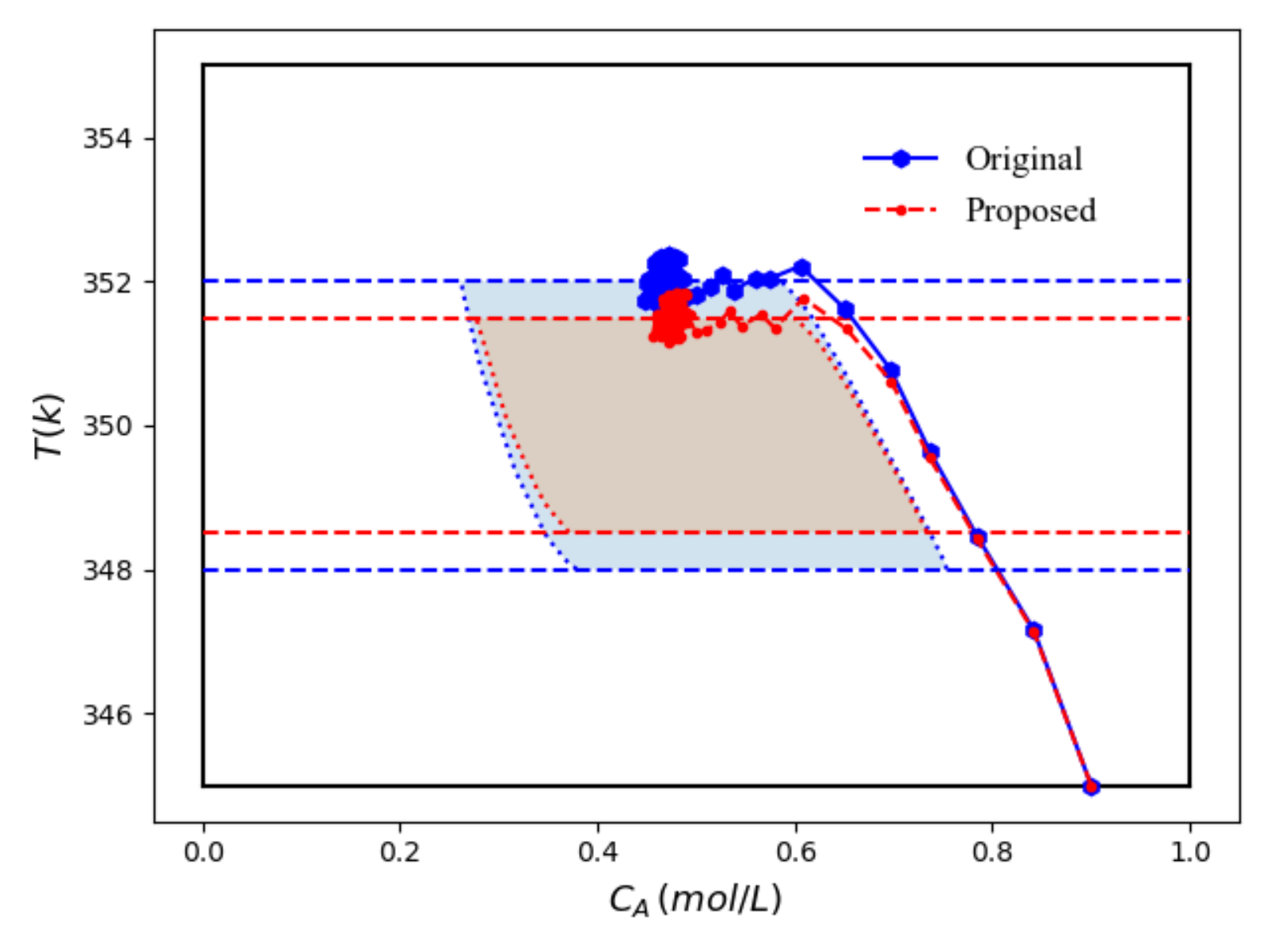}
		\caption{Proposed ZMPC v.s. the nominal ZMPC with economic objective, $x_0 = [0.9,345]$.}
		\label{fig:proposed2}
	\end{figure}
	
	In this section, we validate the performance of the proposed controller. Figure \ref{fig:le0} presents the state trajectories under the proposed ZMPC without any economic objective (\ref{eqn:ZMPC2}) with and without the presence of disturbance. Figures \ref{fig:proposed1} and \ref{fig:proposed2} present the performance of the proposed ZMPC with economic objective (\ref{eqn:ZMPC2_e}) in comparison to the nominal ZMPC based on different initial conditions. In all figures, the solid black box represents the overall state space, and the dashed blue and red lines denote the boundaries of the original target set $\m X_t$ and the modified target set $\tilde{\m X}_t$ respectively. The shaded areas represent the largest CIS inside $\m X_t$ and $\tilde{\m X}_t$. The larger shaded area represents $\m X_t^M$, while the smaller one is $\tilde{\m X}_t^M$.
	
	It can be observed from the blue trajectory in Figure \ref{fig:le0} that without any economic objective, the system converges to a steady-state in the center of the target zone. In the presence of disturbance, as represented by the red trajectory, the state oscillates around the optimal steady state but is able to stay inside the terminal CIS once entered. Thus, for this particular setup, the benefit of the proposed controller is not significant. 
	
	To validate the benefits of the proposed approach, the secondary economic objective is added to the system. With the economic objective considered, the optimal operating condition shifts to the boundary of the target zone. 
	In Figures \ref{fig:proposed1} and \ref{fig:proposed2}, starting from different initial conditions, the state progressions under the control of the nominal ZMPC are shown in blue, while the state trajectories controlled by the proposed ZMPC are red. Under the nominal ZMPC, the state leaves the terminal CIS after converging to the optimal operating point due to the fluctuation caused by the presence of disturbance. These violations on the zone-tracking objective are unfavored and should be eliminated. As presented by the red trajectories, violations are avoided by the proposed ZMPC. Although the state trajectories violate the upper bound of the modified target zone $\tilde{\m X}_t$, they are still enclosed in the actual target zone $\m X_t$. Furthermore, Figures \ref{fig:proposed1} and \ref{fig:proposed2} verify that the proposed controller is able to drive the system state to the same economically-optimal neighborhood even under different initial conditions. 
	
	\subsection{The significance of modifying the target set}
	\label{target_set}
	\begin{figure}[t]
		\centering
		\includegraphics[width=0.8\columnwidth]{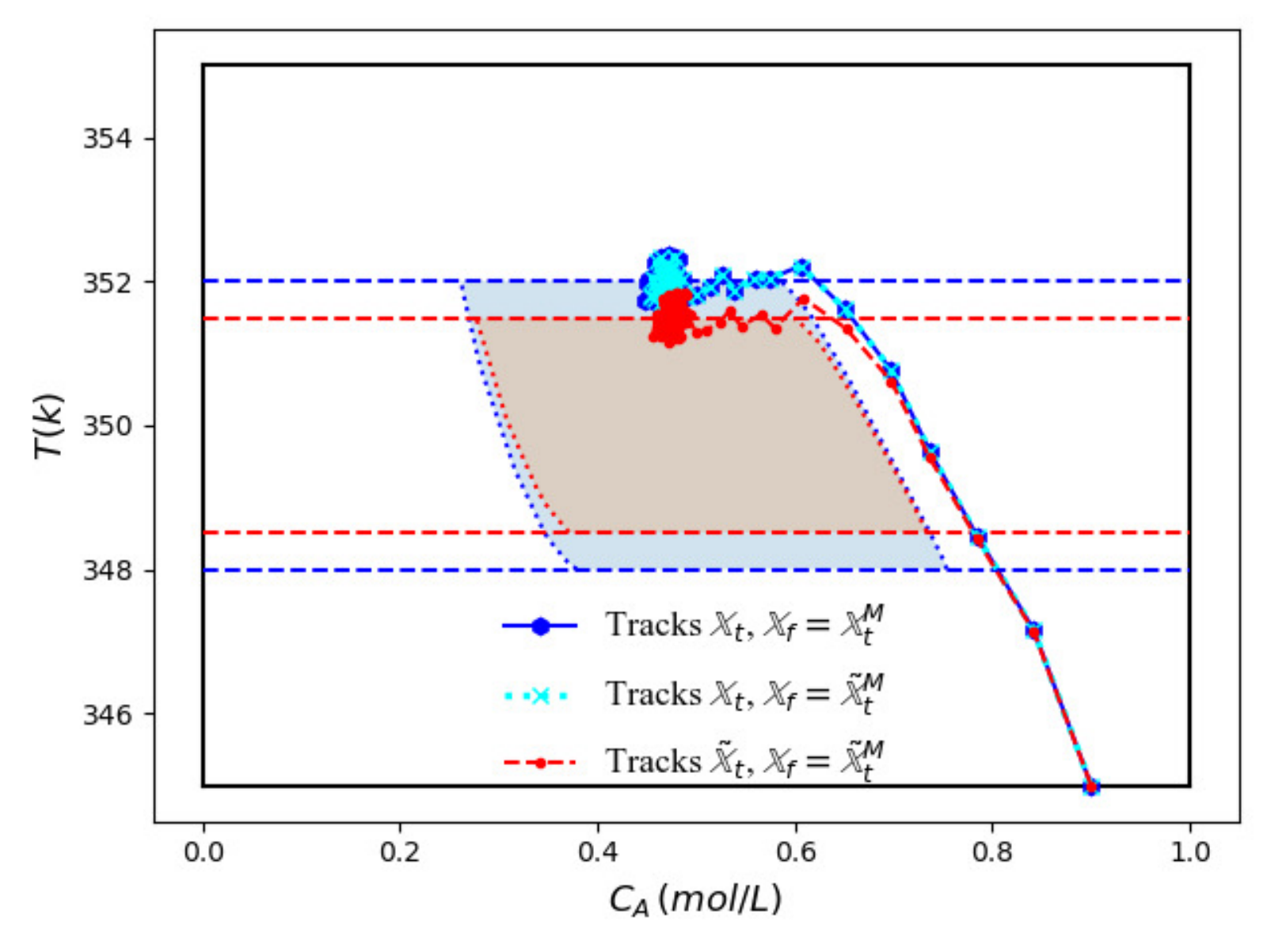}
		\caption{ZMPC with economic objective that tracks the original target set ($\m X_t$) with modified terminal set ($\m X_f = \m X_t^M$).} 
		\label{fig:ShrunkCIS}
	\end{figure}
	One natural question that rises is that instead of modifying just the terminal set, why both the tracking target set and the terminal set are modified in the proposed approach. Figure \ref{fig:ShrunkCIS} investigates the controller performance if the original target set $\m X_t$ is tracked and only the terminal set is modified. The state trajectories under the control of the nominal ZMPC (blue solid trajectory with hexagon markers), the proposed ZMPC (red dashed trajectory with dot markers), and the ZMPC that tracks the original target set $\m X_t$ and the modified terminal set $\tilde{\m X}_t^M$ (cyan dotted trajectory with cross markers) are presented, respectively. It can be observed that if only the terminal set is modified, the optimal state trajectory is identical to that obtained based on the nominal ZMPC. This implies that the control performance is not affected if only the terminal set is modified. This matches the theory proposed in \cite{liu_economic_2018}, which indicates the controller performance remains equivalent in terms of stability as far as $\tilde{\m X}_f \subseteq \m X_t^M$. Note that the state trajectory obtained based on the proposed ZMPC is presented in the figure for reference, which verifies the effectiveness of modifying the tracking target set. 
	
	\subsection{The significance of the terminal constraint}
	\label{stab_validation}
	\begin{figure}[t]
		\centering
		\includegraphics[width=0.8\columnwidth]{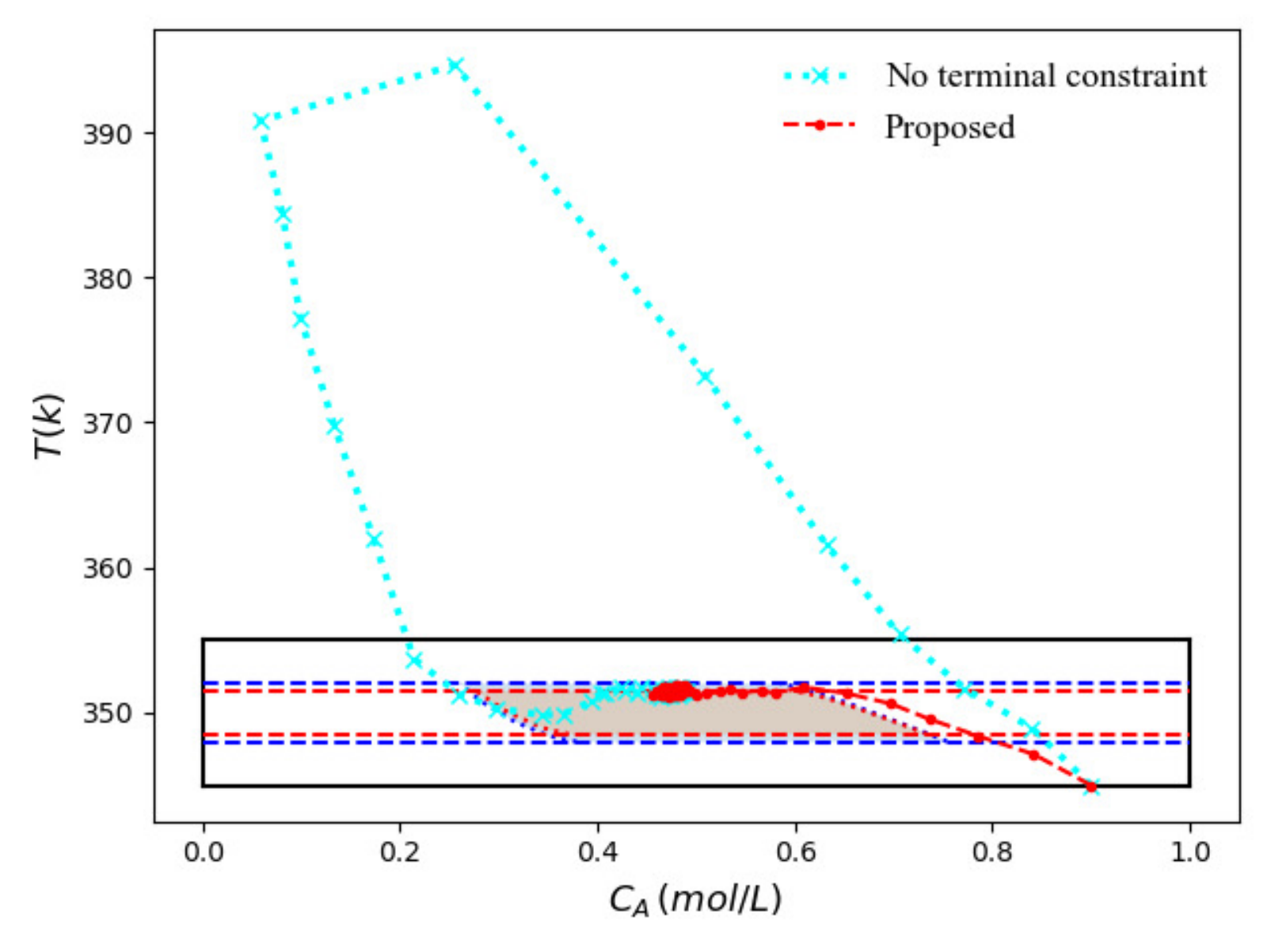}
		\caption{ZMPC without the terminal constraint v.s. the proposed ZMPC with economic objective.}
		\label{fig:term_set}
	\end{figure}
	The significance of the terminal constraint in ensuring closed-loop stability is investigated in this section. With the control horizon $N = 3$, Figure \ref{fig:term_set} shows the state trajectory under the ZMPC without any terminal constraint in cyan with cross markers, while the trajectory obtained based on the proposed controller is presented in red with dot markers for reference. Without the terminal constraint, the controller is able to drive the system to the optimal operating point, however with aggressive violations of the hard state constraint represented by the solid black box. These violations are avoided by the proposed controller with the terminal constraint. Furthermore, it can be observed that under the controller without terminal constraint, the system takes significantly more steps to converge compared to that taken based on the proposed approach, indicating the effectiveness of the terminal constraint in ensuring closed-loop stability and enhancing convergence. 
	
	
	\subsection{The impact of $\gamma$}
	\label{gamma}
	\begin{figure}[t]
		\centering
		\includegraphics[width=0.8\columnwidth]{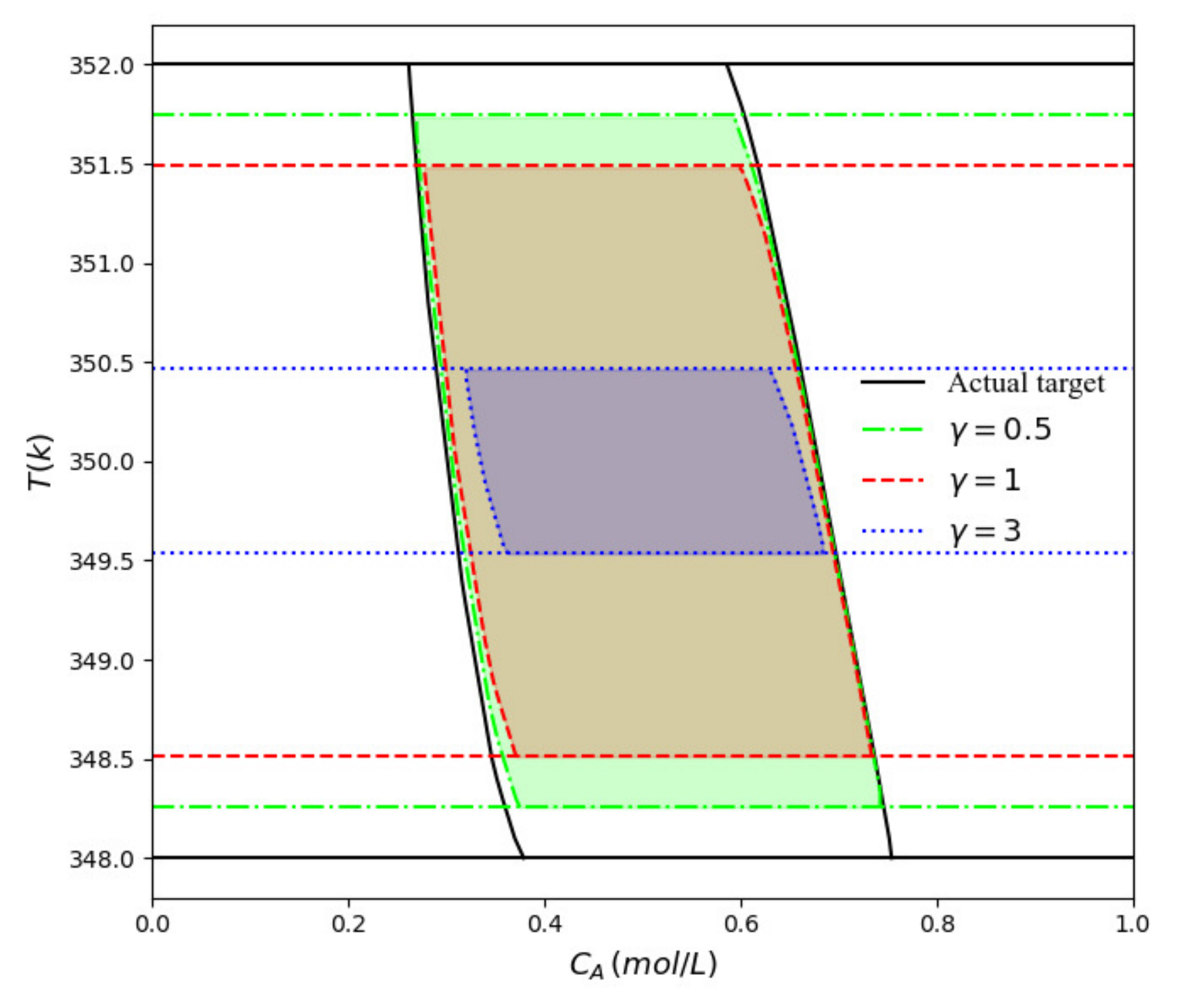}
		\caption{The modified target sets and the corresponding maximum CISs under different $\gamma$.}
		\label{fig:gamma_set}
	\end{figure}
	
	\begin{figure}[t]
		\centering
		\includegraphics[width=0.8\columnwidth]{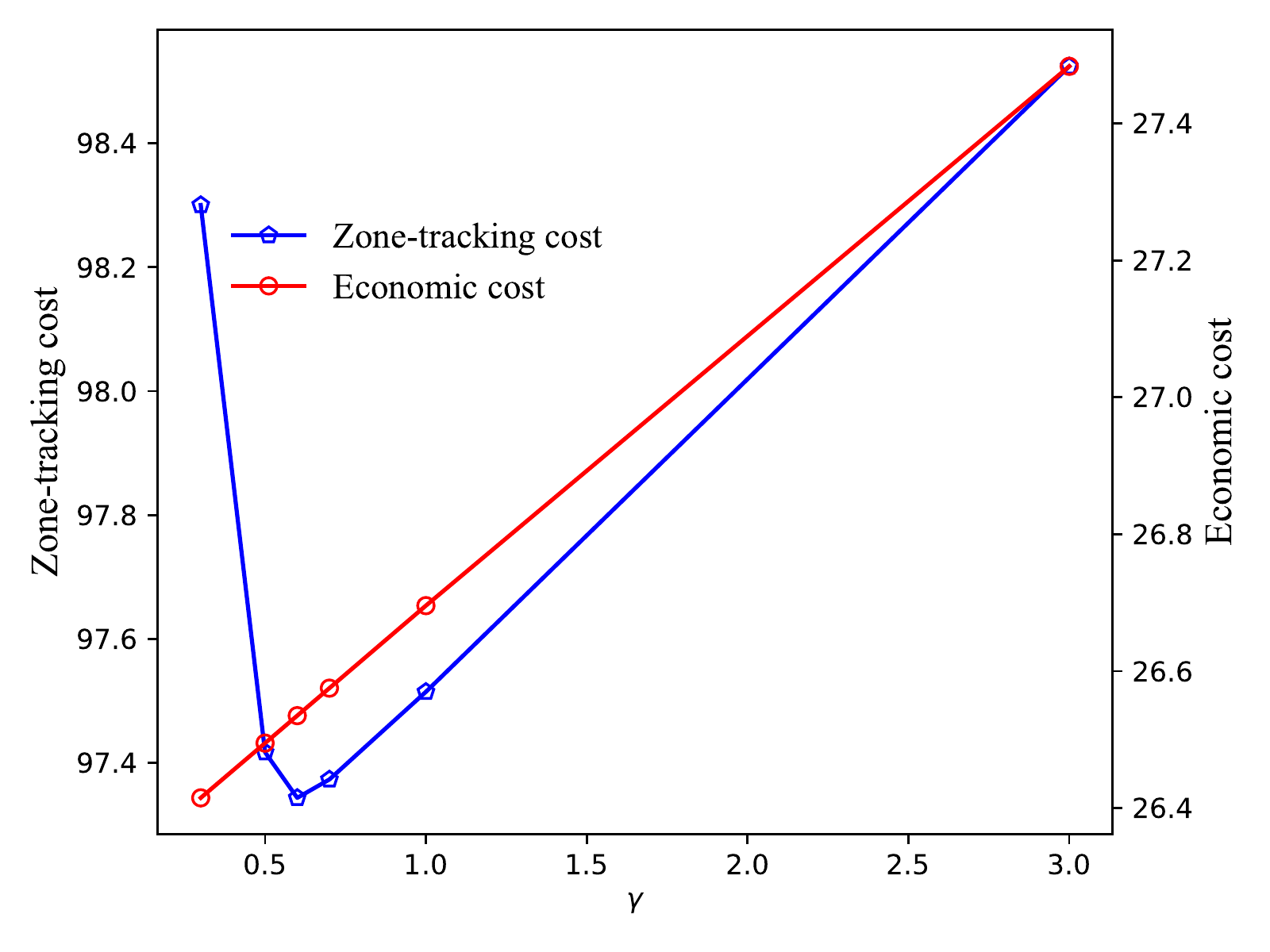}
		\caption{The zone-tracking performance and economic performance under different $\gamma$.}
		\label{fig:gamma_performance}
	\end{figure}
	
	The impact of the tuning parameter $\gamma$ is investigated in this section. Figure \ref{fig:gamma_set} presents the modified target sets and the corresponding maximum CIS inside the modified target sets with different values of $\gamma$. Figure \ref{fig:gamma_performance} displays the accumulated zone-tracking cost and the economic cost as $\gamma$ increases. The accumulated zone-tracking costs are presented in blue with pentagon markers, while the accumulated economic costs are shown in red with circle markers. Note that the accumulated zone-tracking cost is calculated based on the actual target set for a fair comparison, as the modified target zone provided to the controller changes with different $\gamma$. Table \ref{tab:gamma_vio} acts as a supplement of Figure \ref{fig:gamma_performance} and presents the number of violations with respect to the actual target set after the system converges to the optimal operating neighborhood for the first time, plus the average magnitude of violation out of all the violations with different $\gamma$.  
        \begin{table}[t]
		\small
		\centering
		\caption{The number of violations and the average violation value after the first convergence to the actual target set with different $\gamma$.}	
		\renewcommand\arraystretch{2}
		\label{tab:gamma_vio}
		\tabcolsep 15pt
		
		\begin{tabular*}{0.4\textwidth}{ccc}\hline
			$\gamma$ & \# of Vio. & Avg. Vio.\\ \hline
			0.3	&	9	&	0.097 \\
			0.5	&	4	&	0.045 \\
			0.6	&	2	&	0.027 \\
			0.7	&	0	&	- \\
			1	&	0	&	- \\
			3	&	0	&	- \\\hline
		\end{tabular*}
	\end{table}	
	
	The actual target set is bounded by solid black lines in Figure \ref{fig:gamma_set}. The sets with $\gamma$ equal to 0.5, 1, and 3 are bounded by green dash-dotted, red dashed, and blue dotted curves, respectively. 
	The information regarding the optimal operating condition is not considered, both the upper bound and the lower bound of the target zone are shrunk equally. Figure \ref{fig:gamma_set} indicates the size of the target set and the corresponding largest terminal set available to the controller reduce as the value of $\gamma$ increases. This implies the controller becomes more conservative with a smaller feasible region, which leads to a sacrifice in the economic performance. The economic performance is affected in two ways. First, since a smaller feasible zone leads to a shifting in the feasible economically-optimal operating condition. Second, with a more conservative target and terminal zone, the controller tends to drive the system into the zone more aggressively and potentially sacrifice the transit economic performance. This can be seen from Figure \ref{fig:gamma_performance} that the accumulated economic cost function continuously increases as $\gamma$ increases. 
	
	It is however noteworthy that a more conservative controller does not necessarily leads to better zone-tracking performance. It is observed that the accumulated zone-tracking cost first decreases and then increases after reaching a minimum at $\gamma = 0.6$. From Table \ref{tab:gamma_vio}, it can be observed that at $\gamma = 0.6$, two violations are observed with a very small magnitude of the average violation. At $\gamma = 0.7$, no violation is observed after convergence, however, the accumulative zone-tracking cost is slightly higher than that obtained with $\gamma = 0.6$. The reason for the increase in zone-tracking cost as $\gamma$ increase is due to the shrinkage in the feasible operating range and the more aggressive control action chosen by the controller, which sacrifices the transit performance of the system. A larger $\gamma$ helps to reduce the zone-violation and thus the accumulative tracking cost; however, this reduction is saturated once the violations are insignificant ($\gamma = 0.6$) or are fully eliminated ($\gamma = 0.7$). The sacrifice in the transit performance on the other hand, continuously increases as $\gamma$ increase, which slowly overtakes the cost reduced by shrinking the target zone. 
	
	\begin{remark}
		It is observed from Figure \ref{fig:gamma_performance} and Table \ref{tab:gamma_vio} that the optimal zone-tracking performance is not achieved at $\gamma = 1$. This indicates that directly shrinking the target zone by the $x^d_{max}$ only provides a conservative estimation of the amount of shrinkage required. It is essential to tune $\gamma$ to obtain a better zone-tracking performance. On the other hand, it is noteworthy that the economic performance always reduces as $\gamma$ increases. Thus, it is recommended to use the smallest $\gamma$ that provides reasonable zone-tracking performance in applications. 
	\end{remark}

	\section{Conclusions}
	\label{section:conclusion}
	In this work, we proposed a robust ZMPC formulation with guaranteed convergence to the actual target set in the presence of bounded disturbance. This is achieved by modifying the actual target set, which helps to reject the effect of the disturbance. A terminal constraint is utilized to ensure the closed-loop stability of the system. The system state is forced to converge to a CIS inside the modified target set at the end of the control horizon. Without assuming the existence of an optimal steady-state operating condition, this generalized approach provides more degrees of freedom. Furthermore, the proposed design is able to handle a secondary economic objective without affecting closed-loop stability. Apart from the theoretical stability proof, a practical guideline for determining the modified target set is provided. The proposed design is applied to a CSTR system and is proved to be effective from various perspectives. The effect of the tuning parameter in the proposed practical guideline is investigated with the rule of thumb for parameter selection provided.

\section{Acknowledgement}

Financial support from Natural Sciences and Engineering Research Council of Canada is gratefully acknowledged.

\end{document}